# Querying Attack-Fault-Defense Trees: Property Specification in Smart Grid and Aerospace Case Studies

Reza Soltani, University of Twente

Stefano M. Nicoletti, University of Twente

Milan Lopuhaä-Zwakenberg, University of Twente

Mariëlle Stoelinga, University of Twente and Radboud University

## SUMMARY & CONCLUSIONS

This paper introduces AFDL, a logic-based framework for reasoning about safety, security, and defense interactions in Attack-Fault-Defense Trees, which is a model that captures all safety, security, and defense domains in a single framework. We showcase both AFDL and propose a structured domain specific query language, LangAFDL, which enables domain experts to express complex analysis goals through intuitive templates. LangAFDL supports both Boolean and quantified queries as well as minimal cut set analysis, capturing the interplay between safety, security, and defensive measures. We illustrate the expressiveness and utility of the approach through representative queries over two different real-world case studies: Gridshield and Ground Segment as a Service. The formalization lays the automated safety-security groundwork for analyses in mission-critical systems and paves the way for future tool development and integration into design workflows.

## 1  INTRODUCTION

### 1.1  Attack-Fault-Defense Trees (AFDTs)

The increasingly growing complexity of cyber-physical systems led to growth in the risk they face to their continued functioning. These systems must be able to withstand both the accidental failure that can come from its many high-tech components and also (cyber)attacks by malicious actors. Therefore, we need to be able to design defense mechanisms against such safety and security risks. To realize this goal and to build robust designs and resilient cyber-physical systems in general, it is crucial to have a consistent framework for evaluating the safety, security, and impact of countermeasures that specialists from a wide range of fields can understand.

Tree-based risk models, such as Fault Trees (FTs) for safety and Attack Trees (ATs) for security, are often used as frameworks for allowing communication across disciplines. These are hierarchical diagrams consisting of a Top Level Event (TLE) denoting system disruption, and leaves denoting atomic events (component failures and basic attacker actions). There are intermediate events in between, the activation of which is dependent on the type (AND/OR) and activation of their children. People from various disciplines may easily understand these models, and each expert can contribute to the section of the tree that pertains to their area of expertise.

As FTs and ATs solely address safety and security respectively, more complex models are required. To cover the multitude of risks and their countermeasures faced by complex systems, Attack-Fault Trees (AFTs) [1], which can represent dual safety-security hazards, and Attack-Defense Trees (ADTs) [2], that models security risks and their mitigating measures are introduced. Nevertheless, none of these models can adequately capture the security, safety, and countermeasures interplay.

In our previous work [3], we introduced Attack-Fault-Defense Trees (AFDTs), a framework that combines elements of attack trees, fault trees, and defense strategies for capturing safety, security, and countermeasures in a single model.

### 1.2  Querying Fault and Attack Trees

In line with the modeling language extension operated with AFDTs, we adopt the same approach to extend previous querying frameworks developed for fault and attack trees only or for combinations that do not account for defense mechanisms. Very recent work addressed the issue of developing tailored languages that enable practitioners to specify flexible properties on FTs and ATs by proposing three different logics tailored to FTs and ATs, accompanied by model checking algorithms that can check the truth value of formulae.

*Boolean Fault tree Logic.* Previous work [4] proposed a *Boolean Fault tree Logic* (BFL) with which practitioners can: 1. set evidence to analyze what-if scenarios, e.g., what are the Minimal Cut sets (MCSs), given that Basic Event (BE) $A$ or subsystem $B$ has failed? What are the Minimal Path sets (MPSs) given that $A$ or $B$ have not failed? 2. check whether two FT elements are independent or if they share a child that can influence their status. 3. check whether the failure of one (or more) element $E$ always leads to the failure of the TLE. 4. set upper/lower boundaries for failed elements, e.g., would element

*E* always fail if at most/at least two out of *A*, *B* and *C* were to fail? Moreover, if a property does not hold, the BFL framework generates counterexamples, to show why the property fails.

*Probabilistic Fault tree Logic.* Extending the previous framework, [5] presents a *Probabilistic Fault tree Logic* (PFL) to further enhance quantitative analysis capabilities, as probabilities are the prime quantitative metric on FTs. With PFL, one: 1. can check whether the probability of a given element (potentially conditioned by another one) respects a certain threshold, 2. can set the value of one BE in complex formulae to an arbitrary probability value, 3. can check if two BEs or intermediate events are stochastically independent, 4. can also return probability values for given formulae, possibly mapping single elements to an arbitrary probability value. Furthermore, [5] presents *LangPFL*, a domain specific language for PFL that propels the usability of this framework, allowing easier property specification on FTs.

*A Logic for Attack Tree Metrics.* Concerning ATs, [6] develops a *Logic for Attack Tree Metrics* (ATM) to specify a variety of quantitative security properties on these models: the authors present a general framework that considers *security metrics*, such as the "cost" of an attack, "probability" of getting attacked and "skill" of a malicious actor. With ATM, one: 1. can reason about successful/unsuccessful attacks; 2. can check whether metrics, such as the attack cost, are bounded by a given value on single attacks; 3. can compute metrics for a class of attacks and 4. perform quantification over all possible attacks. Note that because ATM uses a general algebraic framework, it allows for the analysis of many different metrics [7].

Finally, [8] combines these frameworks for joint property specification on FT/ATs. However, all these works do not account for the flexibility and the extended expressivity proposed by AFDTs, nor do they consider the pivotal role of defense mechanisms in the overall risk evaluation of a system under analysis.

### 1.3 Our Contribution

In this work, we integrate BFL, PFL, and ATM to enable joint property specification on AFDTs. Our approach grounds the qualitative analysis of AFDTs through reasoning about dependencies between attacks, faults, and defenses. This framework can be further extended for quantitative evaluations for risk assessment and system reliability analysis. Our approach allows us to ask critical safety-security questions, such as: "Is a cyberattack alone sufficient to cause system failure, or do additional faults need to occur?", "What is the probability of system failure if a specific component is compromised?" and "Does a successful attack always require exploiting misconfigurations?". By jointly leveraging these logics, we provide a structured and expressive framework for analyzing and improving the cyber-physical systems' resilience.

## 2 ATTACK-FAULT-DEFENSE LOGIC

In Attack-Fault-Defense Logic (AFDL), we analyze the interplay between attacks, faults, and defenses in AFDTs. The goal is to formally evaluate how failures and attacks lead to system compromise and how defensive mechanisms mitigate these risks. A crucial aspect of AFDL is defining risk scenarios, which involve both attacks and faults. We introduce a risk vector $R$ that explicitly quantifies over attacks and faults while distinguishing them from defenses. This allows us to express queries that explore system behavior under all possible attack-fault scenarios, ensuring that defensive measures are consistently accounted for.

**Definition 1:** We define $R$ as a risk state vector:
$$R = A \cup F \cup D$$
Where $A$ represents a subset of the set of attack steps, $F$ represents a subset of the set of component failures, and $D$ represents a subset of the set of defense steps. Here, attack steps and component failures are semantically distinct: the former represent adversarial actions, while the latter denote accidental malfunctions; defenses mitigate either or both. This distinction allows us to reason about all possible attack and fault combinations while keeping defenses as separate parameters.

Attack-Fault-Defense Logic (AFDL) is a formal framework designed to qualitatively analyze the interplay between cyber-attacks, system failures, and defense mechanisms within AFDTs. AFDL provides a structured approach to reasoning about system vulnerabilities and protective measures. This logic enables analysts to assess the conditions under which a system can fail systematically, the effectiveness of countermeasures, and the necessary conditions for an attack to succeed. AFDL has three categories of queries to qualitatively analyze AFDTs, namely:

- **Boolean Queries (BQ):** This query is in $(R, T \models \varphi)$ format and it evaluates whether a given formula $\varphi$ holds under a specific risk scenario $R$ in the AFDT $T$, returning either false or true.
- **Quantification Queries (QQ):** These queries analyze structural dependencies between BASs, BCFs, and BDSs. They determine whether certain conditions are necessary or sufficient for a specific event to occur. Quantification queries can be in either $T \models \forall_R(\varphi)$ or $T \models \exists_R(\varphi)$ format, where the symbols $\forall_R$ and $\exists_R$ indicate that a property holds for all or at least one risk configuration, respectively.
- **Satisfaction Set Queries (SSQ):** This query type focuses on identifying MRSs, which represent the smallest sets of failures, attacks, and/or defenses whose occurrence leads to reaching a node in the AFDT. These queries can be defined as $[\![MRS(\varphi)]\!]$ where the set of Minimal Risk Scenarios (MRSs) that lead to a particular outcome ($\varphi$) is returned. By analyzing all minimal configurations of attacks, faults, and defenses that satisfy $\varphi$, this category provides deeper insight into the most critical failure-inducing conditions in the system.

AFDL merges BFL, PFL and ATM by syntactically inheriting from these three logics: BQ queries inherit syntax from Layer 1 of BFL [4], PFL [5] and ATM [6], while QQ queries inherit from BFL syntactic Layer 2 and ATM Layer 4, respectively.

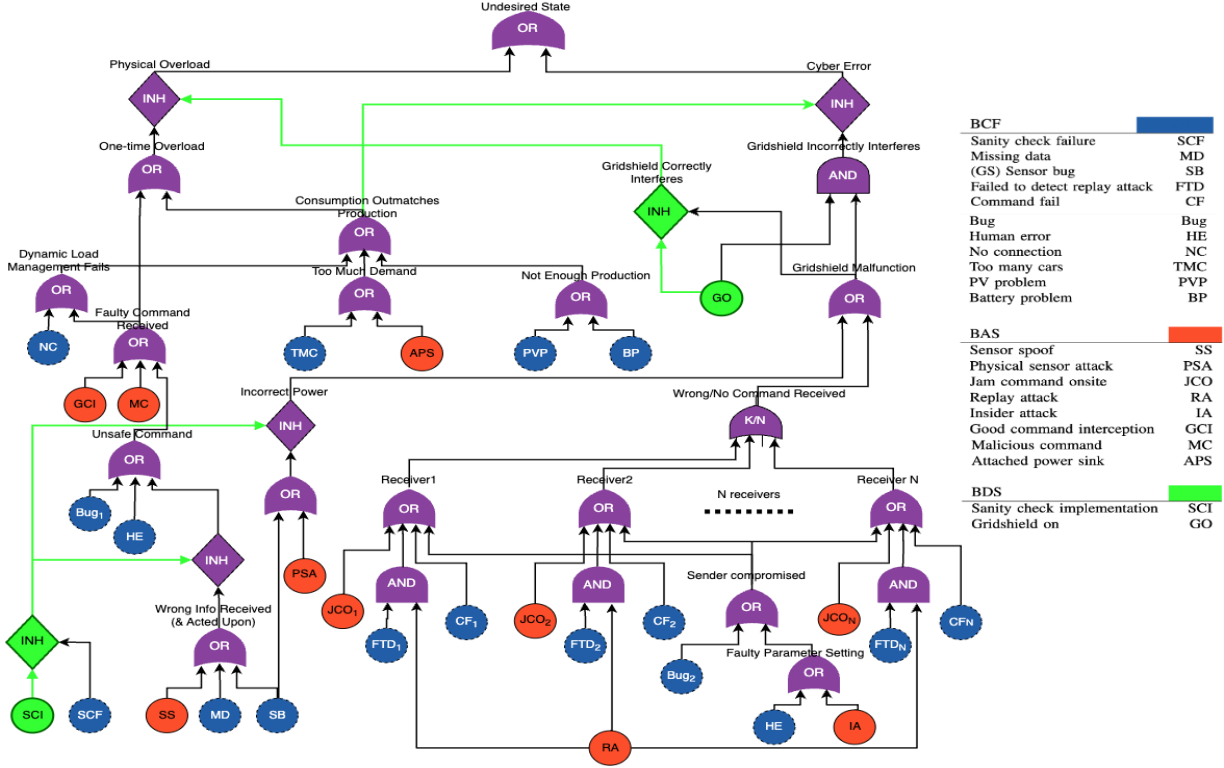

*Figure 1 – Attack-Fault-Defense Tree for Gridshield. BCF, BAS, and BDS stand for Basic Component Failure, Basic Attack Step, and Basic Defense Step, respectively.*

Finally, SSQ queries inherit from all the above frameworks.

## 3 CASE STUDY 1: GRIDSHIELD

### 3.1 Gridshield AFDT description

This case study deploys AFDTs to assess the security and reliability of the Gridshield infrastructure for charging smart electric vehicles (EVs), as presented in [9]. Gridshield is a cyber-physical defense mechanism designed to prevent power grid disruptions by dynamically adjusting EV charging power. It operates through a sender module that monitors the smart grid's critical assets and broadcasts control messages. This mechanism allows the embedded receiver modules in chargers to regulate power consumption. Gridshield complements existing protections, such as Dynamic Load Management Systems (DLMS), which manage charging loads but may fail due to cyberattacks, sensor errors, or cloud service disruptions.

The AFDT model of Gridshield was developed through expert consultations and highlights risks in both DLMS and the Gridshield system. DLMS failures may result from cyber threats aiming to manipulate electricity markets or sensor errors that lead to incorrect load balancing decisions. On the other hand, Gridshield faces vulnerabilities in both sender and receiver modules. Receiver failures occur when incorrect or no commands are received, while sender failures can propagate system-wide due to compromised sensor data or cyber intrusions. Additionally, a critical failure scenario arises when a sufficient number of receiver modules fails, exceeding the system's tolerance threshold. *Figure 1* illustrates the AFDT for Gridshield. We refer the reader to [9] for more details.

The AFDT-based analysis of Gridshield informs practical design choices, reinforcing the importance of combining Gridshield with existing DLMS for enhanced reliability. While the model does not assign specific risk probabilities, it qualitatively demonstrates that large-scale attacks are unlikely without insider involvement. This structured evaluation provides a basis for designing security measures and testing protocols to ensure robust smart grid operations.

### 3.2 Property specification (AFDL)

Having introduced our first case study example, we will now focus our attention on property specification. In this section we will showcase some queries that one can formalize for the AFDT in *Figure 1*. We will do so by presenting statements and their corresponding formulation in our domain specific query language – LangAFDL – starting with properties in natural language and the respective translation in AFDL:

1) Is it possible for an attacker to cause a Cyber Error (CE) without having the Gridshield (GO) activated given that there is a Human Error (HE)?

$$\exists_R(CE[GO \mapsto 0, HE \mapsto 1]) \quad (1)$$

2) Is at least a Jam Command Onsite (JCO) attack sufficient to make the entire smart grid fail (reaching the Undesired State (US)) in the presence of an active Gridshield (GO)?

$$\forall_R((VOT_{\geq 1}(JCO_1, JCO_2, \ldots, JCO_N) \Rightarrow US)[GO \mapsto 1]) \quad (2)$$

3) Is the activation of the Gridshield (GO) always sufficient to prevent Physical Overload (PO) attacks?
$$\forall_R(\neg PO[GO \mapsto 1]) \quad (3)$$
4) Identify all minimal risk scenarios leading to the Undesired State (US) involving the Physical Sensor Attack (PSA), given that the Sanity Check Implementation (SCI) and Gridshield (GO) defenses are not active?
$$[\![MRS(US)[PSA \mapsto 1, SCI \mapsto 0, GO \mapsto 0]]\!] \quad (4)$$
5) Is the given risk scenario minimal with respect to the TLE? And does it cause the TLE to fail if the Gridshield is activated, even if the scenario is not minimal?
$$MRS(US) \wedge US[GO \mapsto 1] \quad (5)$$

Note that we use the single square brackets in all queries to set the value of specific elements in an AFDT (i.e., whether an attack step is taken, a component failure has happened, or a defense is activated). Additionally, we can ask whether a property holds for at least one/for all the possible risk scenarios via quantifiers (∃ and ∀ respectively) as shown in queries (1), (2), and (3). This flexibility enables risk assessors to ask non-trivial questions on the risks their systems are under, from the perspectives of safety (e.g., where components fail) – as done in query (3) – security (e.g., what attackers can achieve) – as per queries (2), (4), and (5) – or both – as in query (1), without overlooking the paramount role of defense mechanisms.

### 3.3 Property specification (LangAFDL)

To ease usability, we showcase how these queries would be specified using LangAFDL, our proposed domain specific language for AFDL. LangAFDL is based on previous domain specific languages exclusive for FT and AT analysis, i.e., LangPFL [5] and LangATM [4,8] and on their seminal composition for property specification on fault trees with attacks [8]. In this work, we extend these frameworks to account for the greater expressivity of AFDTs. Inheriting from these DSLs, LangAFDL is based on structured templates. LangAFDL expresses only a fragment of AFDL: most notably, nesting of formulae is disallowed. By doing so, we retain most of the expressiveness of AFDL while making property specification easier. One can specify assumptions on the status of AFDT elements by utilizing the *assume* keyword. These assumptions will be appropriately integrated in the translated AFDL query: e.g., *set* is used to set the status of AFDT elements – them being attack steps, component failures or defenses – and is translated with the according operators to set evidence, while other more involved assumptions will be the antecedent of an implication. A second keyword separates specified queries from the assumptions and dictates the desired result: *computeall* computes and return desired values, i.e., lists of MRSs, while *check* establishes if a specified property holds. Moreover, we support the use of *decorators* – introduced in [8] – i.e., constructs employed to specify different sets of assumptions. We demonstrate their usage below, in (10). First, we will showcase translations of all the queries in (1)-(5):

*assume:* (6)
    set GO = 0; set HE = 1

*check:*
    exists CE
*assume:* (7)
    set Vot[JCO₁, JCO₂, …, JCO_N] ≥ 1;
    set GO = 1
*check:*
    forall US
*assume:* (8)
    set GO = 1
*check:*
    forall not PO
*assume:* (9)
    set PSA = 1; set SCI = 0; set GO = 0
*computeall:*
    MRS(US)
*assume:* (10)
    *@A1:*    set GO = 1
*check:*
    MRS[US] and *@A1*(not US)

In (10) we can see decorators at work. *@A1* (in orange) is a decorator containing a specific set of assumptions: these are declared under the *assume* keyword. Under the *check* keyword, we specify assumptions for each part of the formula that we want to check: in (10), no assumptions are applied to the minimal risk scenario check – MRS[US] – but assumptions declared in *@A1* – setting the Gridshield defense to active – are applied to *US*. This allows practitioners to clearly grasp the specific what-if scenarios under which various components of their queries are examined. The assumption scope is, however, not restricted in (8): enforcing the Gridshield to be the only active defense mechanism is a global assumption, valid for the entirety of the query under the *check* keyword. Thus, there is no further need for decorators. Properties (6) to (9) showcase the use of quantifiers, ranging over risk scenarios. In these four queries, we can appreciate how different assumptions declared under the *assume* keyword are translated into AFDL queries: e.g., for (6) what-if scenarios is set in AFDL with the usual single square brackets. These technical details are conveniently hidden from the end user who adopts LangAFDL, whose attention should only be invested in defining relevant assumptions without worrying about correct placement for parsing purposes in the lower-level AFDL queries.

## 4 CASE STUDY 2: GROUND SEGMENT AS A SERVICE

### 4.1 GSaaS AFDT description

This case study focuses on the safety and security assessment of the Ground Segment as a Service (GSaaS) model, specifically the cloud-based solution developed by Ascentio Technologies S.A., an Argentinean technological company in the aerospace sector, as presented in [10]. Traditionally, satellite ground segments require dedicated infrastructure that is tightly integrated with proprietary systems. In contrast, Ascentio's GSaaS offers a scalable, on-demand alternative, enabling satellite operators to access ground stations via cloud-native interfaces. This model significantly reduces operational costs and enhances flexibility but also introduces new challenges related to cybersecurity, system reliability, and operational resilience.

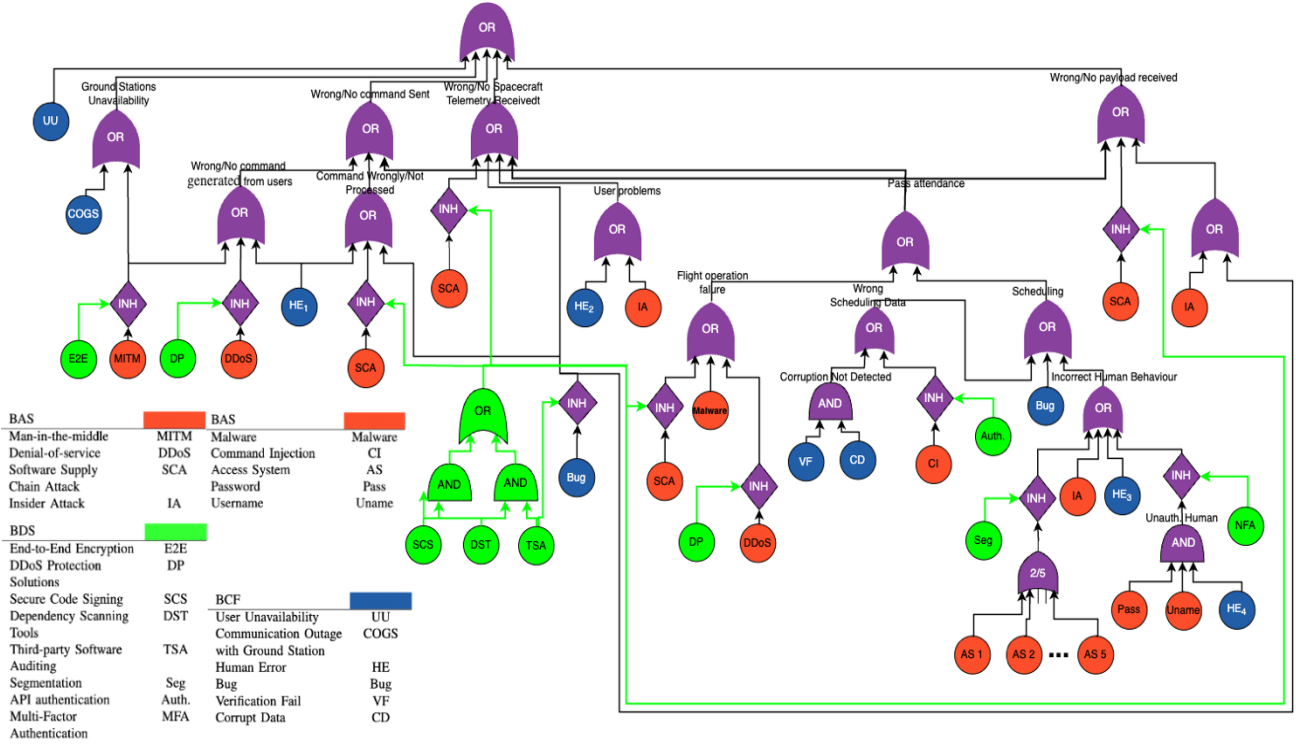

*Figure 2 – Attack-Fault-Defense Tree for the Ground Segment as a Service (GSaaS) model. BCF, BAS, and BDS stand for Basic Component Failure, Basic Attack Step, and Basic Defense Step, respectively.*

We apply AFDT to systematically evaluate safety and security risks in GSaaS, identifying potential vulnerabilities such as cyberattacks, service disruptions, and data breaches. The analysis considers critical failure pathways, including faulty telemetry data or commands that compromise system integrity, procedural or human errors leading to mission failures, and ground station unavailability due to connectivity issues or hardware malfunctions. The AFDT framework provides a structured representation of how these risks propagate through the system and how existing defense mechanisms mitigate them. *Figure 2* illustrates an AFDT for the Ground Segment as a Service (GSaaS) model. We refer the reader to [10] for more details.

By applying AFDT to Ascentio's GSaaS, this case study not only contributes to strengthening its security posture but also provides a methodological blueprint for evaluating similar cloud-based ground segment architectures. The insights gained help refine risk mitigation strategies, ensuring reliable and secure satellite operations while addressing emerging challenges in GSaaS deployments.

### 4.2 *Property specification (AFDL)*

As done in *Section 2.2*, we will showcase meaningful queries in both AFDL and LangAFDL, with a focus on the case study presented above (*Figure 2*). Let us start once again with AFDL properties:

1) Identify all minimal risk scenarios leading to a Scheduling Failure (SF) when both Insider Attack (IA) and Human Error (HE) happen, assuming all defenses are inactive.

$$[\![MRS(SF)[IA \mapsto 1, HE \mapsto 1, E2E \mapsto 0, DP \mapsto 0, SCS \mapsto 0, DST \mapsto 0, TCA \mapsto 0, Seg \mapsto 0, Auth \mapsto 0, MFA \mapsto 0]]\!] \quad (11)$$

2) Is it possible for an attacker to cause a Flight Operation Failure (FOF) without exploiting a Software Supply Chain Attack (SCA), assuming that Dependency Scanning Tools (DST) are inactive?

$$\exists_R (FOF[SCA \mapsto 0, DST \mapsto 0]) \quad (12)$$

3) Are at least two Human Errors (HE) always sufficient to cause a Faulty Payload Reception (FPR), even when all defenses are active?

$$\forall_R ((VOT_{\geq 2}(HE_1, HE_2, HE_3, HE_4) \Rightarrow FPR)[E2E \mapsto 1, DP \mapsto 1, SCS \mapsto 1, DST \mapsto 1, TCA \mapsto 1, Seg \mapsto 1, Auth \mapsto 1, MFA \mapsto 1]) \quad (13)$$

In properties (11), (12), and (13), we examine different aspects of the interaction between attack steps, component failures, and defense mechanisms in the GSaaS system. The property (11) helps in understanding the worst-case scenarios where risk scenarios can lead to scheduling disruptions, highlighting the minimal combinations of attack steps, component failures, and defense mechanisms responsible for such failures. In (12), the query assesses the possibility of attack paths leading to FOF, analyzing whether attackers can still compromise flight operations without targeting software supply chains. In (13), the universal query evaluates the resilience of the system against human errors, even under optimal defense configurations. If the condition holds, it indicates that certain human-induced failures remain unavoidable despite the presence of all defensive mechanisms, emphasizing the critical role of human reliability in ensuring mission success, and guiding a potential redesign. Note that in AFDL we make use of the double square brackets to request the computation of all these satisfying scenarios.

### 4.3 *Property specification (LangAFDL)*

As done in *Section 2.3*, we showcase how an end user could specify properties (11), (12), and (13) only by interacting with LangAFDL. Following, we present respective translations of these queries to our DSL:

*assume:* (14)
    set IA = 1; set HE = 1; set E2E = 0;
    set DP = 0; set SCS = 0; set DST = 0;
    set TCA = 0; set Seg = 0; set Auth = 0;
    set MFA = 0
*computeall:*
    MRS (SF)

*assume:* (15)
    set SCA = 0; set DST = 0
*check:*
    exists FOF

*assume:* (16)
    set Vot[$HE_1$, $HE_2$, $HE_3$, $HE_4$] ≥ 2;
    set E2E = 1; set DP = 1; set SCS = 1;
    set DST = 1; set TCA = 1; set Seg = 1;
    set Auth = 1; set MFA = 0
*check:*
    forall FPR

Queries (14), (15) and (16) showcase the ease of use provided by LangAFDL even when specified properties addressing the complex interactions between safety and security: our DSL handles assumptions placement depending on their complexity and employing the AFDL set evidence operator for translated logic queries like (11) and (12), while also using an implication when translating back to queries like (13). Notably, when we require to compute all minimal risk scenarios – such as for query (11) – we see the introduction of a dedicated LangAFDL keyword, i.e., the *computeall* command in query (14).

## 5 FUTURE WORK

Future work includes extending AFDL with quantitative analysis capabilities, developing model checking algorithms to automatically evaluate queries over AFDTs, and building a practical tool to support the full LangAFDL pipeline from modeling to query evaluation.

### REFERENCES

1. R. Kumar, M. Stoelinga, "Quantitative Security and Safety Analysis with Attack-Fault Trees," Proc. IEEE 18th Int. Symp. High Assurance Systems Engineering (HASE), 2017, pp. 25–32.
2. B. Kordy, S. Mauw, S. Radomirović, P. Schweitzer, "Attack–Defense Trees," J. Logic and Computation, vol. 24, no. 1, 2014, pp. 55–87.
3. R. Soltani, M. Lopuhaä-Zwakenberg, M. Stoelinga, "Safety-Security Analysis via Attack-Fault-Defense Trees: Semantics and Cut Set Metrics," Proc. SAFECOMP 2024, Lecture Notes in Computer Science, vol. 14988, Springer, Cham, 2024.
4. S. M. Nicoletti, E. M. Hahn, M. Stoelinga, "BFL: a Logic to Reason about Fault Trees," Proc. DSN, IEEE/EUCA, 2022, pp. 441–452.
5. S. M. Nicoletti, M. Lopuhaä-Zwakenberg, E. M. Hahn, M. Stoelinga, "PFL: A Probabilistic Logic for Fault Trees," Proc. 25th Int. Symp. Formal Methods (FM), Springer Nature, 2023, pp. 199–221.
6. S. M. Nicoletti, M. Lopuhaä-Zwakenberg, E. M. Hahn, M. Stoelinga, "ATM: a Logic for Quantitative Security Properties on Attack Trees," Under review, SEFM 2023.
7. M. Lopuhaä-Zwakenberg, C. E. Budde, M. Stoelinga, "Efficient and Generic Algorithms for Quantitative Attack Tree Analysis," IEEE Trans. Dependable and Secure Computing (TDSC), 2023.
8. S. M. Nicoletti, M. Lopuhaä-Zwakenberg, E. M. Hahn, M. Stoelinga, "Querying Fault and Attack Trees: Property Specification on a Water Network," Proc. 2024 Ann. Reliability & Maintainability Symp. (RAMS), Albuquerque, NM, USA, 2024, pp. 1–6, doi: 10.1109/RAMS51492.2024.10457796.
9. R. Soltani, B. Ozceylan, M. Lopuhaä-Zwakenberg, C. Kolb, G. Hoogsteen, "Safety and Security Dependencies for Gridshield," Proc. 2024 IEEE PES Innovative Smart Grid Technologies Europe (ISGT EUROPE), Dubrovnik, Croatia, 2024, pp. 1–6, doi: 10.1109/ISGTEUROPE62998.2024.10863084.
10. R. Soltani, P. Diale, M. Lopuhaä-Zwakenberg, M. Stoelinga, "Safety and Security Risk Mitigation in Satellite Missions via Attack-Fault-Defense Trees," arXiv preprint 2504.00988, 2025. Available at: https://arxiv.org/abs/2504.00988.

### BIOGRAPHIES

*Reza Soltani, MSc.*
University of Twente
Enschede, Drienerlolaan 5, 7522 NB, The Netherlands

e-mail: r.soltani@utwente.nl

Reza Soltani is a PhD candidate at the University of Twente (NL) working on integrating safety and security, as well as combined risk analysis. He has contributed to the ERC-funded project CAESAR with the goal of marrying the fields of safety and (cyber)security.

*Stefano M. Nicoletti, Dr.*
University of Twente
Enschede, Drienerlolaan 5, 7522 NB, The Netherlands

e-mail: s.m.nicoletti@utwente.nl

Stefano M. Nicoletti is a postdoc researcher at the University of Twente (NL) working on logics, query languages and model checking algorithms for risk assessment models, such as fault trees and attack trees. He received his PhD from the same university, working on the ERC-funded project CAESAR.

*Milan Lopuhaä-Zwakenberg, Dr.*
University of Twente
Enschede, Drienerlolaan 5, 7522 NB, The Netherlands

e-mail: m.a.lopuhaa@utwente.nl

Milan Lopuhaä-Zwakenberg is an assistant professor at University of Twente (NL), studying safety, security and privacy metrics and their interplay. Before, he was a postdoc at Eindhoven University of Technology (NL) and he received his PhD from Radboud University (NL) on arithmetic geometry.

*Mariëlle Stoelinga, Prof. Dr.*
University of Twente and Radboud University
Enschede, Drienerlolaan 5, 7522 NB, The Netherlands
Nijmegen, Houtlaan 4, 6525 XZ, The Netherlands

e-mail: m.i.a.stoelinga@utwente.nl

Mariëlle Stoelinga is professor of risk management at the Radboud University and the University of Twente (NL). She received a prestigious ERC consolidator grant (CAESAR). She holds an MSc and a PhD degree from Radboud University.